\documentclass[12pt]{article}
\usepackage[colorlinks=true,linkcolor=blue, linktoc=page, urlcolor=blue,citecolor=magenta]{hyperref}
\usepackage{amsmath,amscd}
\usepackage{amsfonts}
\usepackage{amssymb}
\usepackage{graphicx,shortvrb}
\usepackage{epsfig}
\usepackage{newlfont}
\usepackage{cite}
\usepackage[utf8]{inputenc}

\def\cala{\cal A}

\def\caln{{\cal N}}
\def\calo{{\cal O}}

\def\G{\Gamma}

\def\o{\omega}

\def\S{\Sigma}

\begin{document}
\begin{titlepage}

\begin{center}{\huge \bf
A twist on multicenter AdS$_2$ solutions.} 
\end{center}
\vskip .3cm
\vskip 1.5cm

\centerline{\large {{\bf Delaram Mirfendereski$^a$ and Dieter Van den Bleeken$^{a,b}$}}}

\vskip 1.0cm

\begin{center}
$a)$: \sl Primary address\\
 Physics Department, Boğaziçi University\\
 34342 Bebek / Istanbul, TURKEY

\vskip 1cm

$b)$: Secondary address\\
Institute for Theoretical Physics, KU Leuven\\
3001 Leuven, Belgium

\vskip 1cm

\texttt{delaram.mirfendereski@boun.edu.tr\qquad dieter.van@boun.edu.tr}

\end{center}

\vskip 1.3cm \centerline{\bf Abstract} \vskip 0.2cm \noindent

The multicenter solutions of 4d $\caln=2$ supergravity contain a subset of scaling solutions with vanishing total angular momentum. In a near limit those solutions are asymptotically locally AdS$_2\times$ S$^2$, but we show that a higher moment of angular momentum contributes a subtle twist, rotating the S$^2$ with time. This provides some potential hair distinguishing the asymptotics of these scaling solutions from the near horizon geometry of an extremal BPS black hole.

\end{titlepage}

\tableofcontents

\section{Introduction}
The multicentered black hole solutions of 4d $\caln=2$ supergravity \cite{Denef:2000nb,LopesCardoso:2000qm,Bates:2003vx} provide an interesting setting to investigate the BPS spectrum of string theory compactified on a Calabi-Yau manifold and the associated physics problem of black hole entropy and microstates \cite{Gaiotto:2006wm, Denef:2007vg, deBoer:2008zn, Manschot:2014fua, Alexandrov:2018iao}. Through string/M duality these solutions can be lifted to 5 dimensions \cite{Gaiotto:2005gf, Bena:2005ni, Behrndt:2005he, Elvang:2005sa}. It is in this setting that recently a subset of multicenter solutions, often called 'scaling solutions' \cite{Denef:2002ru, Bena:2006kb, Denef:2007vg, Bena:2007qc} have been revisited and their asymptotic AdS$_2$ nature explored \cite{Bena:2018bbd}, see also \cite{Bena:2007ju,Lunin:2015hma}. In this short note we point out that somewhat surprisingly the asymptotic geometry typically has a fibred structure, with an S$^2$ rotating over AdS$_2$. Interestingly this rotation is not linked to the total angular momentum (which for these scaling solutions vanishes, similar to single center black holes) but to a higher moment of the angular momentum. For lack of deeper understanding we call this new feature the 'twist'. This twist provides some hair that distinguishes the asymptotics of AdS$_2$ multicenter solutions from the near horizon black hole AdS$_2\times$S$^2$ geometry. Since it has been argued that it are precisely the scaling solutions that correspond to the exponential majority of black hole microstates \cite{Bena:2012hf, Lee:2012sc, Manschot:2012rx} a precise holographic interpretation of the twist would be highly interesting. We leave this last problem for future work. After reviewing some technicalities of the multicenter solutions in section \ref{sec1} and spelling out some details on both the far and near region of a scaling solution in section \ref{scal} we come to the point in section \ref{fn} and derive the asymptotic AdS$_2$ geometry \eqref{asmet} revealing the subtle presence of the twist \eqref{twistvec}. We end with some comments in section \ref{coms}.

\section{Reminder of $\caln=2$ multicenter black holes}\label{sec1}
The multicenter solutions of $\caln=2$ supergravity are dyonic black holes interacting through electromagnetic and scalar field induced forces in such a way that stable bound states are formed. Although rather intricate, exact explicit solutions are known, for a review see \cite{Denef:2007vg,VandenBleeken:2008tsa}. The theory has $n$ complex scalar fields $t^A$ and $n+1$ $U(1)$ gaugefields $(A^0,A^A)$. For a generic multicenter solution these fields and the metric take the following form:
\begin{eqnarray}
 ds^2 & = & -\frac{1}{\S}(dt+\,\o)^2+\S\, dx^idx^i\,, \label{metric}\\
  A^0 & = & \frac{-L}{\S^2}\left(dt+\o\right)+\omega_0\qquad  A^A = \frac{H^AL-Q^{3/2}y^A}{H^0\S^2}
  \left(dt+\o\right)+{\cala}_\mathrm{d}^A\,,\nonumber\\
 t^A&=&\frac{H^A}{H^0}+\frac{y^A}{Q^{\frac{3}{2}}}\left(i\S-\frac{L}{H^0}\right),\nonumber
\end{eqnarray}
The whole solution is determined in terms of $2n+2$ harmonic functions $H=(H^0,H^A,H_A,H_0)$ which for $N$ dyonic charges $\Gamma_a=(p^0_a,p^A_a,q_A^a,q_0^a)$ at positions $\vec{x}_a$ in the spatial $\mathbb{R}^3$ take the simple form
\begin{equation}
H=\sum_a\frac{\Gamma_a}{r_a}+h\,,
\end{equation}
These harmonic functions enter the fields above through a set of auxiliary functions. First they define the $y^A$, obtained by formal solution of the quadratic equations\footnote{Here the constant symmetric three tensor $D_{ABC}$ and the symplectic inner product $\langle E_1,E_2\rangle=-E^0_1E_0^2+E_1^AE^2_A-E_A^1E_2^A+E_0^1E_2^0$ are those associated to the particular 4d $\caln=2$ supergravity under consideration. In the case of Calabi-Yau compactifications these two objects are naturally determined by the internal geometry. The precise value of the constants $D_{ABC}$ will play however no role in the current paper and all of our discussion hence also applies to situations with no known embedding in string theory, as for example some 'magic' ${\caln}=2$ theories \cite{Gunaydin:1983rk}.  }
\begin{equation}
D_{ABC}y^Ay^B=-2H_CH^0+D_{ABC}H^AH^B\,,
\end{equation}
and then
\begin{equation*}
\begin{split}
 Q^3=(\frac{1}{3}D_{ABC}y^Ay^By^C)^2\,, \qquad& L=H_0(H^0)^2+\frac{1}{3}D_{ABC}H^AH^BH^C-H^AH_AH^0\,,\\
 &\S=\sqrt{\frac{Q^3-L^2}{(H^0)^2}}\,.
\end{split}
\end{equation*}
Furthermore there are the one-forms\footnote{The Hodge star is that of flat $\mathbb{R}^3$.}
\begin{equation}
d{\cala}_\mathrm{d}^A = \star dH^A\,,\quad d\omega_0 =\star dH^0 \,, \qquad
 d\omega=\star\langle dH,H\rangle\label{finom}
\end{equation}
At the technical level the bound state nature of these solution appears through a set of equations restricting the (coordinate) distances $r_{ab}$ between the centers:
\begin{equation}
\sum_{b,\, b \neq a}
   \frac{\langle
    \G_a,\G_b\rangle}{r_{ab}}=\langle h,\G_a\rangle\,.\label{denef eqns}
\end{equation}
The constants $h=(h^0,h^A,h_A,h_0)$ appearing in the harmonic functions set the asymptotic values of the scalar fields\footnote{More precisely it is the $h$ being determined by $t^A_\infty$, in such a way that $\langle h, \sum_a \Gamma_a\rangle=0$ \cite{Denef:2000nb}.} and as such correspond to the choice of vacuum. When these constants are non-zero the multicenter solutions are easily seen to be asymptotically flat. When one approaches one of the centers, $\vec{x}\rightarrow \vec{x}_a$, the geometry becomes (for a generic charge $\Gamma_a$) AdS$_2\times$S$_2$, which one recognizes as the near-horizon geometry of an extremal Reisner-Nordstrom black hole, with the scalars taking constant 'attractor' values \cite{Ferrara:1995ih}. Finally it is important to point out that the solutions are stationary with a total angular momentum given by \cite{Denef:2000nb}
\begin{equation}
\vec{J}=\frac{1}{2}\sum_{a<b}\langle\Gamma_a,\Gamma_b\rangle \hat x_{ab}\,.
\end{equation} 
Note that of course the special case of a single center reproduces a standard extremal BPS black hole \cite{Ferrara:1995ih} without angular momentum.

\section{A far and near limit for scaling solutions}\label{scal}
We should point out that although the $3N$ (coordinate) positions of the dyonic black hole centers are constrained by the $N-1$ equations \eqref{denef eqns}, $2N-2=3N-(N-1)-3_{\mathrm{c.o.m.}}$ remain free, leading to an interesting space of solutions (see \cite{deBoer:2008zn, deBoer:2009un} for some first explorations of these spaces). For a generic set of charges $\Gamma_a$ there will be both a minimal and maximal distance between the centers, but in some special case this is not so. In particular the relative coordinate positions of the centers can be made arbitrary small when the charges $\Gamma_a$ are such that there exist a set of positive numbers $s_{ab}=s_{ba}$, among which each triple satisfies the triangle inequalities and
\begin{equation}
\sum_{b,\,b\neq a}\frac{\langle \Gamma_a,\Gamma_b\rangle}{s_{ab}}=0\,.
\end{equation}
Indeed, it directly follows that then $r_{ab}=\xi s_{ab}$ solves the constraint equations \eqref{denef eqns} in the limit $\xi\rightarrow 0$. The above 'scaling' conditions have not, as far as we are aware, been studied/solved in general (for $N>3$), but one can find example solutions for any number of centers\footnote{Take for example $N$ ordered points $\vec{x}_k\in \mathbb{R}^3$, $k\in \mathbb{Z}_N$. Defining $l_k=|\vec{x}_{k+1}-\vec{x}_k|$ one can choose $\langle\Gamma_a,\Gamma_b\rangle=(b-a) l_a$ when $|a-b|=1$ and zero otherwise. In this case $s_{ab}=|\vec{x}_a-\vec{x}_b|$ shows $\langle\Gamma_a,\Gamma_b\rangle$ satisfy the scaling conditions.}. Clearly if a set of charges $\Gamma_a$ satisfies the scaling conditions then so does an arbitrary overall rescaling of these charges, and one can also freely rescale their positions; hence the name.  

The supergravity solution degenerates in an interesting way when the coordinate positions of the centers approach each other \cite{Bena:2007qc,Denef:2007vg}, in particular the physical distance between the centers does not vanish. To understand more clearly what happens it is useful to consider this limit in a slightly different but equivalent way. As on a technical level one is essentially comparing inverse distances to the constants $h$ in the harmonic functions we can study it from that perspective. Let us introduce a parameter $\lambda$ by redefining $h=\lambda \tilde h$, and consider sending $\lambda\rightarrow 0$ while keeping $\tilde h$ and $\vec{x}_a$ fixed. This procedure produces two different supergravity solutions, depending on how we treat the coordinates $(t,x^i)$ in this limit. We'll refer to these two solutions as the far and near limits respectively.
\subsection{The far limit}
Here we rescale the coordinates via $(t,x^{i})=\lambda^{-1}(\tilde t,\tilde x^i)$ and keep the tilded versions fixed in the $\lambda\rightarrow 0$ limit. As now the original coordinates $x^i\gg x_a^i$ one sees that the new coordinates $\tilde{x}^i$ parametrize the region far away from the charged centers. Additionally, if we would define the rescaled positions $\tilde x^i_a$ they go to zero and so the limit also describes the centers approaching each other. Although the supergravity solution has a rather intricate form involving a number of auxiliary functions a closer look reveals that much of this structure is homogeneous under the above rescaling. If after taking the limit one drops the tildes one finds that the solution has essentially remained intact, the only difference being the replacement
\begin{equation}
H=\sum_a\frac{\Gamma_a}{r_a}+h\quad\rightarrow\quad H=\frac{\sum_a \Gamma_a}{r}+h
\end{equation}
So the far limiting procedure reproduces the single center solution of total charge $\Gamma_\mathrm{t}=\sum_a \Gamma_a$, in particular at large $r$ the constant in the harmonic function dominates and the solution is asymptotically flat. Physically what happens is that the original centers develop a stronger and stronger gravitational warping deep in the center which for an observer far away becomes indistinguishable from a single extremal black hole carrying the total charge while nothing much happens to the asymptotics of the original solution. Note that for this procedure to make sense as a continuous limit one needs to keep track of the constraint equations \eqref{denef eqns}, which reduce to the scaling conditions
\begin{equation}
\sum_{b,\, b \neq a}
   \frac{\langle
    \G_a,\G_b\rangle}{r_{ab}}=0 \label{scalcond}
\end{equation}
Note that an interesting physical consequence of these conditions is that the angular momentum of the solution vanishes:
\begin{equation}
\sum_{b}\frac{\langle\Gamma_a,\Gamma_b\rangle}{r_{ab}}=0\ \ \Rightarrow\ \ 0=\sum_{a,b}\frac{\langle\Gamma_a,\Gamma_b\rangle}{r_{ab}}\vec{x}_a=\sum_{a<b}\frac{\langle\Gamma_a,\Gamma_b\rangle}{r_{ab}}(\vec{x}_a-\vec{x}_b)=2\vec{J}\label{angarg}
\end{equation}
This is of course in agreement with the fact that also the total angular momentum of the corresponding single center black hole is zero.

All this might suggest that when the centers of a multicenter solution approach each other a single centered black hole is obtained. Although this is true for the far region we'll see in the next subsection this not at all the case in a near region.

\subsection{The near limit} In this case we do not rescale the coordinates at all, rather keeping $t,x^i$ fixed as $\lambda\rightarrow 0$. This limit is immediate to perform as it leaves the full solution intact, simply putting $h$ to zero. Apart from imposing the scaling conditions \eqref{scalcond}, it simply amounts to replacing the harmonic functions as
\begin{equation}
H=\sum_{a}\frac{\Gamma_a}{r_a}+h\quad\rightarrow\quad H=\sum_{a}\frac{\Gamma_a}{r_a}
\end{equation}
Contrary to the far limit, in the near limit the solution retains its multi-centered nature as it does not differ from the original near any of the centers, i.e. when $r\rightarrow r_a$. The large $r$ behaviour has however drastically changed as there
\begin{equation}
H=\frac{\sum_a \Gamma_a}{r}+{\calo}(r^{-2})
\end{equation}
This suggests that the near limit at large distances is no longer asymptotically flat but should behave as the near horizon of a single center black hole of total charge $\Gamma_\mathrm{t}=\sum_a \Gamma_a$\,, which is AdS$_2\times$S$^2$. It seems a simple picture emerges where the far distance behavior of the near limit matches perfectly with the near behavior of the far limit, both coinciding with the near horizon geometry of a single extremal black hole. Although this is roughly correct there is a small, but we believe important, twist to this intuitive picture which is the main point of this paper. As we will see the large $r$ behaviour of the near limit does not exactly reproduce the near horizon geometry but rather some hair remains through a spinning of the 2-sphere. Technically this originates from carefully keeping track of the ${\calo}(r^{-2})$ term in the harmonic functions as we will explain in some more detail the next section.
 
\section{Far asymptotics of the near limit}\label{fn}
To go beyond the naive analysis of the large distance behavior of the near limit of scaling solutions made at the end of the previous section we will need to keep track of a subleading term in the harmonic functions: 
\begin{equation}
H=\frac{\Gamma}{r}+\frac{\Delta^i \hat x^i}{r^2}+{\calo}(r^{-3})\,.\label{harmexp}
\end{equation}
Here we introduced the electro-magnetic dipole
\begin{equation}
\Delta^i=\sum_a \Gamma_a x_a^i\,.
\end{equation}
Let us stress that the unexpected twist we'll uncover does not originate in subleading terms in an asymptotic expansion of the metric.  Rather, as we'll see, for certain terms the naive leading part will vanish promoting a 'subleading' part to the dominant contribution.  

For the rest of this section we'll focus on the metric as nothing unexpected happens in the gauge fields or scalar expansions, as can be checked by the reader. The main non-triviality of the metric \eqref{metric} is encoded in the warp factor $\Sigma$. It is readily calculated by inserting \eqref{harmexp} in the auxiliary functions, that its leading behavior at large distance is
\begin{equation}
\Sigma=\frac{S}{4\pi\, r^2}+{\calo}(r^{-3})\,,
\end{equation}
where $S(\Gamma_\mathrm{t})$ \cite{Bates:2003vx, Shmakova:1996nz} is a constant that has the physical interpretation of entropy (or horizon area).  

The key point is now the contribution of $\omega$ \eqref{finom} to \eqref{metric}. As was pointed out in \cite{Denef:2000nb} it generically behaves at large distances as $r^{-1}$, with the coefficient directly proportional to (and responsible for) the total angular momentum of the solution. But since for scaling solutions the total angular momentum necessarily vanishes (see \eqref{angarg}) the term at order $r^{-2}$ becomes the leading contribution. An explicit calculation reveals
\begin{equation}
\omega=\frac{K^i}{r^2}\epsilon_{ijk}\hat x^jd\hat x^k+{\calo}(r^{-3})
\end{equation}
where
\begin{equation}
\vec{K}=\frac{\langle\Gamma,\vec{\Delta} \rangle}{2}=\frac{1}{2}\sum_{a<b} \langle
        \G_a,\G_b\rangle \vec{x}_{ab}\label{twistvec}
\end{equation}
Note that we can always choose coordinates such that $\vec{K}$ is oriented along the z-axis, such that in spherical coordinates the expression for $\omega$ then becomes
\begin{equation}
\omega=-\frac{K}{r^2}\sin^2\theta\, d\phi +\calo(r^{-3})
\end{equation}
If the metric were asymptotically flat, where the  both the time-like and spatial warp factor would go to a constant, this $\calo(r^{-2})$ behavior of $\omega$ would remain some subleading angular momentum multipole effect. But the asymptotics have changed by putting $h=0$. The time-like warp factor now blows up like $r^2$ enhancing the contribution from $\omega$ while at the same time the spatial warp factor falls off like $r^{-2}$ tempering the growth of the spatial sphere, in exactly such a way that both contributions become of the same order.  Writing this out produces the far near metric: 
\begin{equation}
ds^2=-\frac{4\pi\,r^2}{S}dt^2+\frac{S}{4\pi\,r^2}dr^2+\frac{S}{4\pi}\left(d\theta^2+\sin^2\theta(d\phi+A)^2\right)\label{asmet}
\end{equation}
here we recognize an S$^2$ fibered over AdS$_2$, with a flat connection
\begin{equation}
A=\frac{16\pi^2 \,K}{S^2}dt
\end{equation}
So interestingly enough the far near metric is not exactly the near far metric but has the extra twist that the 2-sphere is rotating as time in AdS$_2$ flows, with the rate of rotation set by the intriguing quantity \eqref{twistvec}, which for lack of better current understanding we might just as wel refer to as the 'twist' vector\footnote{We refrain from calling $\vec{K}$ spin as it should be clearly distinguished from the angular momentum $\vec{J}$ that vanishes.}. Note that the twist can apparently be removed by a coordinate transformation $\tilde\phi=\phi+\frac{4\pi K}{S^2}t$.

\section{Comments}\label{coms}
Scaling multicenter solutions provide an interesting source of highly non-trivial asymptotic AdS$_2$ geometries. Recently \cite{Bena:2018bbd} argued why they could provide important new insights and directions to 2d holography and they explored some of the first physical properties and consequences. In this note we showed that these solutions might be even richer than naively expected, in that they retain a subtle extra twist (or hair) that is not present in the empty AdS$_2$ background obtained from the near horizon geometry of an extremal black hole. We end with a number of small comments.
\begin{itemize}
\item First a small note of caution. Although it definitely appears as if the twist provides a leading contribution to the asymptotic metric it might be naive to simply treat on equal footing the $dtd\phi$ and $d\phi^2$ components. Directly related is the question if the asymptotic coordinate transformation $\tilde \phi=\phi+\frac{4\pi K}{S^2}t$, that could remove the twist, is indeed large and physically relevant or not. What exactly the correct asymptotic boundary conditions are, and the corresponding asymptotic symmetries, is a subtle issue that needs further careful analysis. This would require a precise 2d bulk theory containing all the relevant fields, which as far as we are aware has not been previously formulated. A simple sphere reduction keeping the connection $A$ as a 2d gauge field seems to have problems with consistency and so a larger framework might be needed.  In a lift to 5d terms of the order $r\,dtd\phi$ are generated \cite{Bena:2018bbd}, but it is unclear if this has any implications for our discussion in 4d.
\item If we are more optimistic this result has potentially interesting physical consequences. It has been assumed that it is exactly the scaling solutions that are key in understanding black hole entropy \cite{Bena:2012hf, Manschot:2012rx}. On the microscopic side because they seem to be associated with an exponential number of 'pure-Higgs' states and on the gravity side because they closely resemble the black hole. The scaling solutions satisfy the condition \cite{Chowdhury:2015gbk} that the angular momentum of black hole microstates should vanish and it is interesting that  the twist uncovered here provides a new observable that can differentiate, even asymptotically, between different AdS$_2$ scaling solutions. The natural arena to try and understand the twist better and a provide a potential connection to microstates is of course holography. For the moment we have nothing to add to the interesting discussion and list of references in \cite{Bena:2018bbd}, but we hope to investigate this further in the future. 
\item We should point out that also in the AdS$_3\times$S$^2$ limit of multicenter solutions a spinning sphere made its appearance \cite{deBoer:2008fk}. We see however no direct relation with the twist here, as the spin there has a direct interpretation as angular momentum, or R-charge in the dual field theory.
\end{itemize}

\section*{Acknowledgements}
It is a pleasure to thank A. Castro for an interesting discussion and J. Raeymaekers for comments on the manuscript. DM and DVdB are partially supported by T\"UBITAK grant 117F376.

\bibliographystyle{utphys}
\bibliography{twist}

\end{document}